\documentclass[11pt]{article}

\usepackage[T1]{fontenc}
\usepackage[utf8]{inputenc}
\usepackage{mathptmx}
\usepackage[margin=1in]{geometry}
\usepackage{amsmath,amssymb,amsthm}
\usepackage{graphicx}
\usepackage{booktabs}
\usepackage{tabularx}
\usepackage{array}
\usepackage{enumitem}
\usepackage{microtype}
\usepackage{xcolor}
\usepackage[numbers,sort&compress]{natbib}
\usepackage{authblk}
\usepackage[colorlinks=true,linkcolor=blue!50!black,citecolor=blue!50!black,urlcolor=blue!50!black]{hyperref}

\newcommand{\E}{\mathbb{E}}

\newcommand{\Dhist}{\mathcal{D}_{\mathrm{H}}}
\newcommand{\Dcur}{\mathcal{D}_{\mathrm{C}}}
\newtheorem{definition}{Definition}
\setlist{itemsep=2pt,topsep=4pt}

\title{Digital Twins for Clinical Development:\\ A Five-Level Framework from Prognostic Adjustment to In-Silico Trials}
\author{Jingyuan Yang}
\affil{AbbVie}
\date{November 2026}

\begin{document}
\maketitle

\begin{abstract}
Digital twins are increasingly discussed in drug development, often narrowly as virtual control arms or synthetic data. We propose a broader framework for data-driven clinical digital twins, organized into five levels according to how much of the trial data matrix of baseline covariates, treatment, and responses is generated by a statistical or machine-learning model trained on historical, external, or real-world data. The levels are: (1)~prognostic score adjustment, (2)~response trajectory prediction, (3)~virtual control, (4)~synthetic population, and (5)~fully synthetic patients and responses. For each level we describe the estimation workflow, the development decision it can support, and the principal statistical guardrail. As the share of model-generated information increases, the inferential weight placed on the model grows and regulatory maturity generally declines. We argue that credibility requirements should therefore scale with context of use, characterized jointly by model influence and decision consequence, following a risk-based credibility framework. We identify four recurring credibility questions (drift and transportability, estimand alignment, compounding error, and data quality and provenance), summarize the current regulatory landscape, and discuss bottlenecks in data, model transferability, collaboration, and lifecycle governance. The central message for statisticians is to start from the decision rather than the model, and to earn trust use by use.
\end{abstract}

\noindent\textbf{Keywords:} digital twin; prognostic covariate adjustment; virtual control; synthetic data; in-silico trial; matching-adjusted indirect comparison; model credibility; estimands

\vspace{0.5em}
\noindent\textit{This article is based on a talk given at the Bay Area Biotech-Pharma Statistics Workshop (BBSW), November 5, 2026. It does not communicate results of AbbVie-sponsored research.}

\section{Introduction}\label{sec:intro}

The term \emph{digital twin} has moved quickly from engineering into clinical development. In much of the current discussion, however, it is equated with a single application: replacing or augmenting a control arm with model-generated outcomes, or more generally producing ``synthetic data.'' For statisticians this framing is too narrow. It obscures the very different inferential roles that a predictive model can play, from contributing a single covariate to a randomized comparison to generating every patient and outcome in a simulated trial.

The question that matters is not what a model can generate, but which development decision its predictions can responsibly support, and what evidence is needed before relying on it for that decision. This article organizes data-driven clinical digital twins around that question. Our contributions are:
\begin{enumerate}
  \item A five-level framework that classifies applications by which components of the trial data matrix are observed and which are model-generated (Section~\ref{sec:framework}).
  \item For each level, a description of the workflow, the value it offers across development, and the main statistical guardrail (Section~\ref{sec:levels}).
  \item A risk-based view of credibility in which evidentiary requirements scale with model influence and decision consequence, together with four recurring credibility questions (Section~\ref{sec:credibility}).
  \item A summary of the regulatory landscape and of the practical bottlenecks that stand between current practice and a ``living'' clinical digital twin (Sections~\ref{sec:regulatory}--\ref{sec:living}).
\end{enumerate}

\section{Lessons from industrial digital twins}\label{sec:industry}

Digital twins earned their reputation outside medicine by linking a model to a real-world decision. An early precursor is Apollo~13 in 1970: after an oxygen tank exploded en route to the Moon, NASA engineers used ground-based simulators kept in sync with the spacecraft to evaluate the situation and rehearse recovery procedures before the crew carried them out. In 2002, Michael Grieves proposed, within a product-lifecycle framework at the University of Michigan, that every physical product have a virtual counterpart \citep{grieves2017digital}. NASA's 2010 technology roadmap formally defined the digital twin \citep{shafto2010nasa}, and through the 2010s aerospace used streaming engine data for predictive maintenance. Applications now extend to factories, warehouses, cities, and the Earth's climate.

Three elements together make a twin: a \emph{virtual model} representing one specific system; a \emph{live data link} keeping the model in sync with its real counterpart; and a \emph{decision loop} in which predictions change what happens next. A predictive model without the decision loop is not a twin.

Contemporary industry practice can be summarized as \emph{simulate before you build}. Examples showcased at NVIDIA GTC in March 2026 \citep{nvidia2026gtc} include a blueprint that simulates the power, cooling, compute, and networking of an AI data center before construction; fleets of autonomous forklifts trained and tested in a virtual warehouse before deployment; and the combination of physiological simulation and AI toward patient-specific twins for care decisions. The workflow in each case is to simulate, test, and optimize virtually, and to commit physical resources only afterward.

Industrial twins succeed because four ingredients come together: (i)~dense, continuous data from sensors; (ii)~models that combine mechanistic structure with learned patterns; (iii)~sufficient compute to run thousands of scenarios before a decision; and (iv)~a closed validation loop in which every prediction is eventually checked against what actually happened. Clinical development already uses simulation, typically at the group level under distributional assumptions, for example in sample-size calculation and in the evaluation of adaptive designs. The opportunity is to extend this capability toward patient-level trajectories and full trial scenarios. The four ingredients also serve as a gap analysis for pharmaceutical development, to which we return in Section~\ref{sec:bottlenecks}.

\section{Clinical digital twins}\label{sec:clinical}

\begin{definition}
A \emph{clinical digital twin} is a computational representation of a biological system or patient, informed by relevant data, that simulates disease progression or response under specified conditions to support a defined development decision.
\end{definition}

The final clause is essential: the intended decision determines the twin's construction, data requirements, and validation. Twins can be built in three ways. \emph{Mechanistic} twins model biology, physiology, and drug action. \emph{Data-driven} twins learn from clinical trial and real-world data using statistical or machine-learning models. \emph{Hybrid} twins combine biological mechanism with data-driven prediction. We focus on data-driven twins, where biostatisticians are most directly involved.

Table~\ref{tab:uses} illustrates uses across the development life cycle, which vary considerably in maturity.

\begin{table}[!htbp]
\centering
\caption{Illustrative uses of data-driven clinical digital twins across development.}\label{tab:uses}
\begin{tabularx}{\linewidth}{@{}lX@{}}
\toprule
Stage & Illustrative use \\
\midrule
Early development & Explore dose, regimen, and response scenarios \\
Trial and phase planning & Refine eligibility criteria, endpoints, and sample size; inform early go/no-go decisions \\
Confirmatory trials & Improve precision through prognostic adjustment while preserving randomization \\
Post-approval & Support indirect comparisons \\
\bottomrule
\end{tabularx}
\end{table}

\paragraph{Notation.} For patient $i$ let $X_i$ denote baseline covariates, $T_i \in \{0,1\}$ treatment assignment (0 = control), and $Y_i$ the response, which may be a vector of longitudinal measurements, a safety outcome, or a time-to-event outcome. Let $\Dhist = \{(X_j, T_j, Y_j)\}$ denote historical, external, or real-world patient-level data used to train a model, and $\Dcur$ the data from the current trial. We write $Y_i(t)$ for the potential outcome under treatment $t$. Across all levels, a prognostic or generative model is trained and validated on $\Dhist$, locked, and only then applied to the current trial.

\section{A five-level framework}\label{sec:framework}

\begin{figure}[!htbp]
\centering
\includegraphics[width=0.82\linewidth]{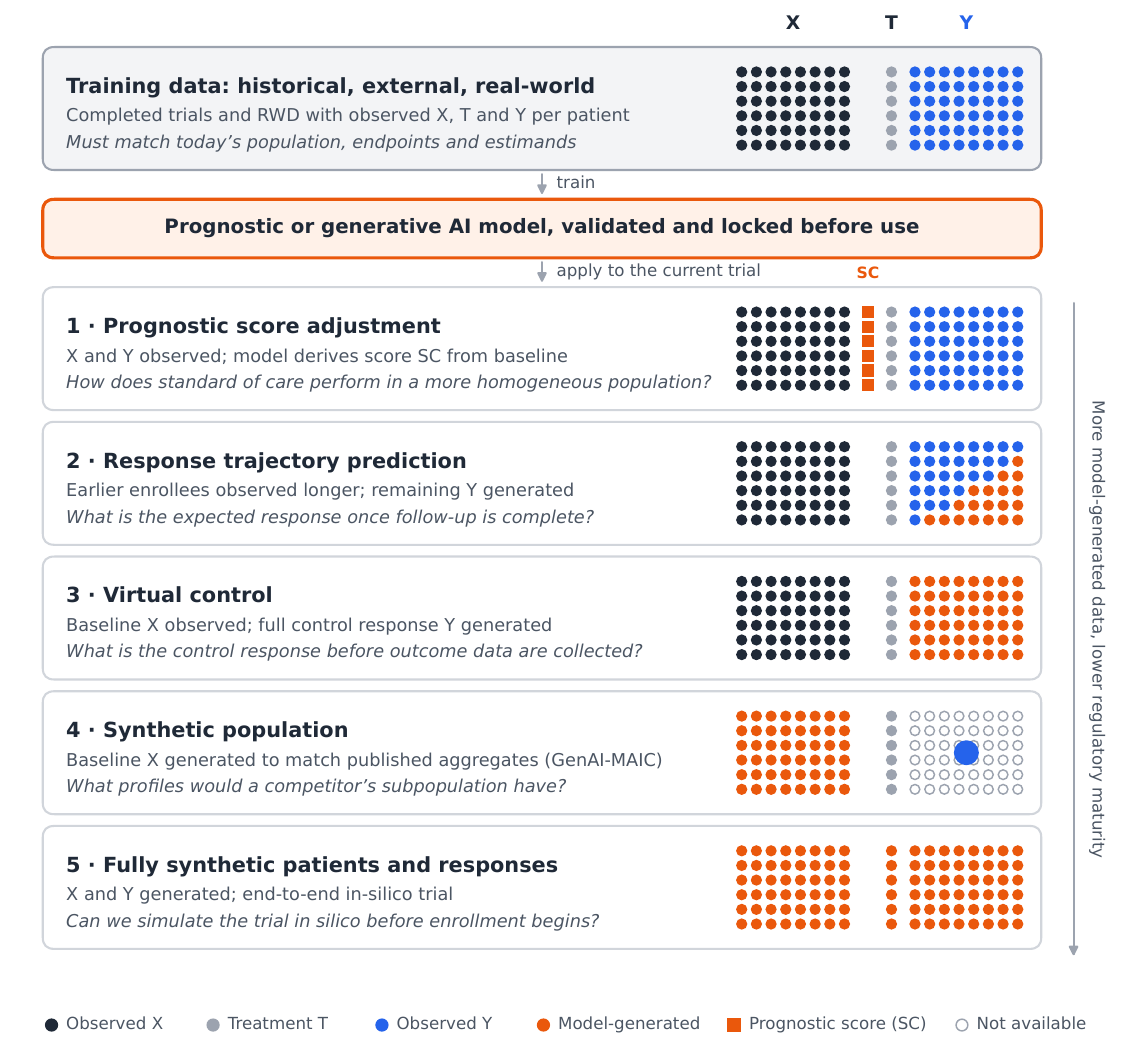}
\caption{Five levels of data-driven clinical digital twins. A model is trained on historical, external, or real-world patient-level data with observed $X$, $T$, and $Y$, validated and locked, and then applied to the current trial. Each row is a patient; orange marks model-generated information. At Level~4 the large blue dot denotes a published aggregate result that stands in for unavailable patient-level outcomes.}\label{fig:levels}
\end{figure}

The five levels differ in which components of the trial data matrix $(X, T, Y)$ are observed and which are generated by the model trained on $\Dhist$. Figure~\ref{fig:levels} depicts each level as a patient-by-variable matrix, and Table~\ref{tab:levels} summarizes the observed and generated components.

\begin{table}[!htbp]
\centering
\caption{Observed and model-generated components at each level.}\label{tab:levels}
\small
\begin{tabularx}{\linewidth}{@{}>{\raggedright\arraybackslash}p{3.3cm}>{\raggedright\arraybackslash}p{2.6cm}>{\raggedright\arraybackslash}p{3.3cm}X@{}}
\toprule
Level & Observed & Model-generated & Question addressed \\
\midrule
1. Prognostic score adjustment & $X$, $T$, $Y$ & One prognostic score per patient, from $X$ & How much more precise can the randomized comparison be? \\
2. Response trajectory prediction & $X$, $T$, early $Y$ & Remaining, not-yet-observed $Y$ & What will the response be once follow-up is complete? \\
3. Virtual control & $X$ (and $Y$ on treatment) & Full control-arm trajectory $Y(0)$ & What is the control response without collecting it? \\
4. Synthetic population & Aggregate summaries of a target population & Individual baseline profiles $X$ & What patient profiles would a comparator subpopulation have? \\
5. Fully synthetic patients and responses & None from the future trial & $X$ and $Y$ & Can the trial be rehearsed in silico before enrollment? \\
\bottomrule
\end{tabularx}
\end{table}

Two trends move together across the levels: the share of generated information increases, and regulatory maturity generally decreases. The levels are distinct use cases, not interchangeable sources of evidence. Two applications may share a model architecture yet carry entirely different inferential weight.

\section{The levels in detail}\label{sec:levels}

\subsection{Level 1: Prognostic score adjustment}

At Level 1 the model contributes exactly one number per patient; baseline covariates and outcomes remain observed. The workflow is:
\begin{enumerate}
  \item \textbf{Train outside the trial.} Using historical control data in $\Dhist$, estimate a prognostic function $m(x) \approx \E[Y \mid X = x, T = 0]$.
  \item \textbf{Score every randomized patient.} Apply the locked model to each patient's baseline covariates, $S_i = \hat m(X_i)$, before any trial outcomes are examined.
  \item \textbf{Adjust the comparison.} Include $S_i$ as a prespecified covariate, for example in the linear model
  \begin{equation}
    Y_i = \beta_0 + \tau T_i + \beta_1 S_i + \gamma^\top X_i + \varepsilon_i .
  \end{equation}
\end{enumerate}
This is covariate adjustment with a well-chosen ``super-covariate'' \citep{schuler2022procova}. Under a linear model with residual correlation $\rho$ between $S$ and $Y$ within arm, the asymptotic variance of the adjusted estimator $\hat\tau$ is approximately $(1-\rho^2)$ times that of the unadjusted difference in means. Because the variance scales with the residual variance divided by sample size, a proportional reduction in residual variance translates roughly into the same proportional reduction in the sample size needed for a given precision. The gain can therefore be used either as higher power at fixed sample size or, when validated gains justify it, as a smaller study. In small Phase~2 trials the score also helps account for chance baseline imbalance.

As an illustration, a retrospective analysis of the AWARE trial in Alzheimer's disease reported approximately 9--15\% lower residual variance across selected cognitive outcomes after adjustment for a digital-twin prognostic covariate \citep{wang_aware}. Such results indicate potential rather than a guarantee for a future trial.

\paragraph{Guardrail.} Randomization and observed outcomes anchor the inference. Provided the score is derived from baseline data alone and prespecified, a poorly predictive score costs efficiency but not validity of the randomization-based comparison. This property explains why Level~1 is the most mature level, supported by the EMA qualification opinion on PROCOVA \citep{ema2022procova} and consistent with FDA guidance on covariate adjustment \citep{fda2023covariate}.

\subsection{Level 2: Response trajectory prediction}

At Level 2, baseline and available follow-up are observed, and the model forecasts the remainder of each patient's trajectory. Partition $Y_i = (Y_i^{\mathrm{obs}}, Y_i^{\mathrm{mis}})$ at a data cut. A longitudinal model trained on mature data in $\Dhist$ provides a predictive distribution
\begin{equation}
  p\bigl(Y_i^{\mathrm{mis}} \mid X_i, T_i, Y_i^{\mathrm{obs}}\bigr).
\end{equation}
Earlier enrollees have more observed visits; later enrollees have more to be predicted. Drawing $M$ completed datasets from this distribution and computing the planned estimator on each yields a distribution of plausible final estimates, from which one can assess, for example, the probability that the eventual estimate would cross a prespecified go/no-go threshold.

The value lies in earlier decision scenarios for internal go/no-go and next-study planning, rather than waiting until follow-up is complete.

\paragraph{Guardrail.} Forecasts are scenarios, not data. Uncertainty should be propagated through multiple draws, in the spirit of multiple imputation; the model should be back-tested at earlier data cuts against outcomes that subsequently became available; and blinding and interim-analysis rules must remain intact.

\subsection{Level 3: Virtual control}

At Level 3, the model generates each enrolled patient's full response trajectory under control, $\hat Y_i(0) = \hat f(X_i)$, where $\hat f$ is trained on relevant historical control data. For the same baseline profile we then hold two different kinds of information: the treatment response $Y_i(1)$, measured in the current trial, and a modeled control response $\hat Y_i(0)$. There are two uses with very different risk.

\emph{Augmentation} supplements concurrent controls with modeled control outcomes, so that fewer concurrent controls are required. \emph{Replacement} omits the concurrent control arm altogether and estimates the effect among treated patients by, for example,
\begin{equation}
  \hat\tau_{\mathrm{VC}} = \frac{1}{n_1}\sum_{i:\,T_i=1}\bigl\{Y_i - \hat f(X_i)\bigr\}.
\end{equation}
The bias of this estimator is $\E\bigl[\hat f(X) - \E\{Y(0)\mid X\} \mid T=1\bigr]$ in the current trial population. Good predictive performance on held-out historical data does not guarantee that this quantity is small in the current trial, because population drift, changes in standard of care, or differences in endpoint ascertainment can shift $\E\{Y(0)\mid X\}$. Without a concurrent control, the bias is not identifiable from the trial data and must be addressed through design and sensitivity analysis. Transportability of the training data, endpoint alignment, and sensitivity to model bias must therefore be examined \citep{burman2024}.

\paragraph{Guardrail.} A generated control is not an observed counterfactual. Predictive accuracy alone cannot establish a treatment effect without a valid study design and explicit causal assumptions.

\subsection{Level 4: Synthetic population}

At Level 4, the model generates baseline profiles $X$ for a target population, while outcomes come from a separate source, typically published aggregate results. Suppose only summary data are available from a comparator trial, such as marginal means and proportions $\bar x^\ast$. The workflow is:
\begin{enumerate}
  \item \textbf{Define the target.} Specify the comparator trial or clinically relevant subgroup to emulate.
  \item \textbf{Generate candidate profiles.} Sample individual baseline profiles $\tilde X$ from a covariate distribution learned on historical trials in a similar indication, constrained to match the published marginal summaries.
  \item \textbf{Filter internal individual patient data.} Select internal patients who resemble the synthetic population, varying the unknown correlations among covariates as a sensitivity analysis, and compare outcomes in the matched cohort with the external summaries.
\end{enumerate}
Standard matching-adjusted indirect comparison (MAIC) reweights observed internal patients with weights $w_i = \exp(\alpha^\top X_i)$ chosen so that $\sum_i w_i (X_i - \bar x^\ast) = 0$ \citep{signorovitch2010maic,jiang2024}. We refer to the generative variant described here as \emph{GenAI-MAIC}, an exploratory extension that uses generated profiles to help construct the comparison population. Its potential advantage is that a learned joint distribution can represent plausible covariate dependence when only marginal summaries are published.

\paragraph{Guardrail.} Generating baseline profiles does not recover a comparator's patient-level outcomes. The value is in informing indirect comparison and planning, not in replacing randomized head-to-head evidence.

\subsection{Level 5: Fully synthetic patients and responses}

At Level 5, both baseline profiles and responses are generated, giving an end-to-end in-silico trial. Table~\ref{tab:level5} lists the components.

\begin{table}[!htbp]
\centering
\caption{Components of an in-silico trial at Level 5.}\label{tab:level5}
\begin{tabularx}{\linewidth}{@{}lX@{}}
\toprule
Trial component & Simulation role \\
\midrule
Population & Generate eligible baseline profiles $\tilde X \sim \hat g$ \\
Treatment and conduct & Specify allocation, visit schedule, adherence, and dropout assumptions \\
Responses & Generate outcomes $\tilde Y \sim \hat f(\cdot \mid \tilde X, T)$ \\
Analysis & Repeat the planned analysis across $B$ simulated trials \\
\bottomrule
\end{tabularx}
\end{table}

This generalizes the clinical trial simulation statisticians already perform, replacing parametric distributional assumptions with patient-level generative models. Across $B$ replicates, teams can estimate operating characteristics such as power, type~I error, and bias under alternative protocols, comparing eligibility criteria, sample size, endpoint timing, and treatment scenarios before committing patients and resources. Generative approaches such as TWIN-GPT point in this direction \citep{twingpt2024}, and model-informed simulation already has an established place in regulatory science \citep{fdaMS}.

\paragraph{Guardrail.} The rehearsal is only as good as its assumptions. Errors in $\hat g$ and $\hat f$ compound, so the generators should be validated against real data and the conclusions tested for sensitivity to misspecification. Level~5 is most useful today for early exploration and Phase~2--3 design; regulatory use is context-specific.

\section{Credibility scaled to context of use}\label{sec:credibility}

\begin{figure}[!htbp]
\centering
\includegraphics[width=0.8\linewidth]{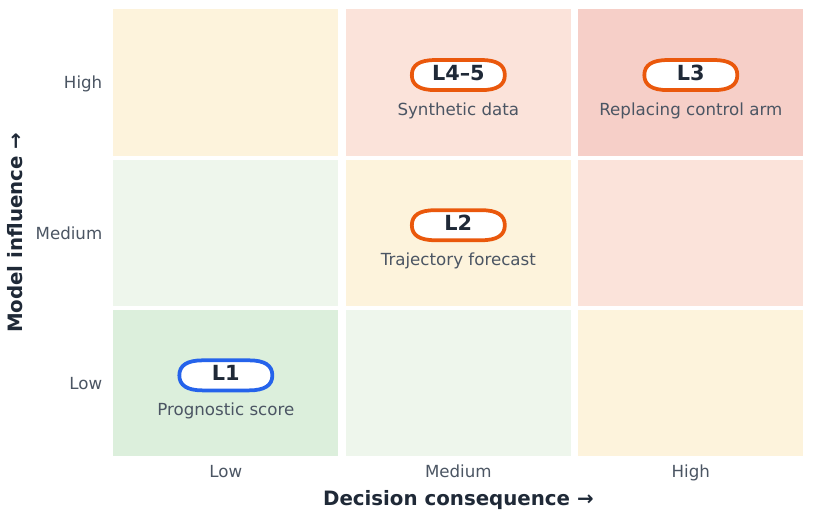}
\caption{Illustrative positioning of the five levels by model influence and decision consequence, adapted from the FDA risk-based credibility framework. Darker cells indicate a higher credibility burden. Positions depend on context of use; the same model moves to the right as its use moves toward a regulatory conclusion.}\label{fig:credibility}
\end{figure}

The evidence a twin requires depends on how much a decision relies on the model and how costly an incorrect decision would be, not simply on how much data the model generates. Adapting the risk-based credibility framework in FDA's draft guidance on AI \citep{fda2025ai}, we characterize each context of use by two dimensions: \emph{model influence} and \emph{decision consequence} (Figure~\ref{fig:credibility}).

Level~1 has low model influence because randomization and observed outcomes anchor inference. Replacing a concurrent control arm with Level~3 virtual controls to support a regulatory conclusion places substantial weight on modeled responses and lies at the high end of both dimensions. Levels~2, 4, and 5 are most naturally exploratory planning tools, in which the sponsor uses scenarios to inform internal decisions.

Four questions recur at every level:
\begin{enumerate}
  \item \textbf{Drift and transportability.} Does today's population, including its standard of care, resemble the training data? Historical controls can become outdated quickly when background therapy changes.
  \item \textbf{Estimand alignment.} Does the analysis address the intended ICH~E9(R1) estimand \citep{ich2019e9r1}? It is necessary to understand how intercurrent events were handled in the training data and whether that matches the strategy in the current trial. A model trained on on-treatment data cannot implicitly supply a treatment-policy estimand.
  \item \textbf{Compounding error.} When generated outcomes depend on generated baseline profiles, as at Levels~4 and~5, errors propagate and may multiply through the workflow. Uncertainty quantification must carry through every generative step.
  \item \textbf{Data quality and provenance.} Are the data traceable, harmonized, and fit for the context of use?
\end{enumerate}
These questions apply whether the model informs an internal go/no-go decision or supports a regulatory submission; what changes is the required strength of the answers.

\section{Regulatory landscape}\label{sec:regulatory}

Regulatory precedent is strongest at Level~1 and thins as model generation increases. Table~\ref{tab:reg} lists key milestones.

\begin{table}[!htbp]
\centering
\caption{Selected regulatory milestones relevant to clinical digital twins.}\label{tab:reg}
\begin{tabularx}{\linewidth}{@{}llX@{}}
\toprule
Year & Body & Milestone \\
\midrule
2022 & EMA & Qualification opinion for PROCOVA: prognostic score adjustment in Phase~2/3 trials with continuous outcomes \citep{ema2022procova} \\
2023 & FDA & Final guidance on adjusting for covariates in randomized clinical trials \citep{fda2023covariate} \\
2025 & FDA & Draft guidance on AI to support regulatory decision-making, with a credibility framework tied to context of use \citep{fda2025ai} \\
2026 & FDA, EMA & Joint guiding principles of good AI practice in drug development \citep{fdaema2026principles}; an HMA--EMA group is formulating AI guidance for clinical development \\
\bottomrule
\end{tabularx}
\end{table}

The 2022 and 2023 milestones directly support Level~1. The 2025 and 2026 documents do not endorse any particular digital-twin application; rather, they provide a vocabulary and a process for asking what evidence a given use requires. The recurring message is to define the context of use precisely, build credibility systematically, and engage regulators early.

\section{Bottlenecks}\label{sec:bottlenecks}

Returning to the four industrial ingredients, the main gaps in pharmaceutical development lie in data and validation rather than compute.
\begin{enumerate}
  \item \textbf{Longitudinal data fit for the question.} Trial and real-world data must be linked with aligned endpoints, clear provenance, and appropriate access. A large historical database can shrink substantially once population drift and relevance to the intended use are taken into account.
  \item \textbf{Models that transfer to new settings.} Calibration should be tested across populations, not only discrimination within one, and uncertainty in treatment-response assumptions should be quantified.
  \item \textbf{Secure collaboration across datasets.} Shared standards and federated approaches may allow sponsors to learn across data sources without exchanging patient-level data, provided that privacy and leakage checks remain in place.
  \item \textbf{A repeatable clinical workflow.} Model ownership, version control, independent testing, and rules for updates must be defined. Such lifecycle management sustains credibility after initial validation.
\end{enumerate}

Compute, by contrast, is scaling rapidly. In January 2026, NVIDIA and Eli Lilly announced a joint investment of up to \$1 billion over five years in talent, infrastructure, and compute \citep{lilly2026nvidia}; in March 2026, Roche announced combined on-premises and cloud infrastructure of more than 3{,}500 GPUs \citep{roche2026gpu}. For clinical twins, such capacity supports training richer models, running trial scenarios in parallel, and quantifying uncertainty at scale. It makes the workflow feasible but does not establish clinical validity.

\section{Toward a living digital twin}\label{sec:living}

The long-term opportunity is not a one-time simulation but a repeatable loop linking predictions, decisions, and observations. Table~\ref{tab:mapping} maps industrial capabilities to clinical counterparts, each subject to validation.

\begin{table}[!htbp]
\centering
\caption{Industrial capabilities and their clinical counterparts.}\label{tab:mapping}
\begin{tabularx}{\linewidth}{@{}XX@{}}
\toprule
Industrial capability & Clinical counterpart \\
\midrule
Simulate before committing to a design & Rehearse protocols before enrollment \\
Test behavior across many operating scenarios & Stress-test decision rules and trial assumptions \\
Combine mechanistic models with learned patterns & Model patient trajectories and trial operations \\
Refresh models with real-world observations & Update forecasts as relevant clinical data arrive \\
Check predictions against operating performance & Compare predictions with outcomes and revalidate \\
\bottomrule
\end{tabularx}
\end{table}

A living twin earns trust through a cycle that repeats over the model's lifecycle, consistent with the FDA draft guidance and the FDA--EMA principles \citep{fda2025ai,fdaema2026principles}:
\begin{enumerate}
  \item \textbf{Define the use and risk:} the decision being made, the model's influence on it, and the consequences of error.
  \item \textbf{Validate for that use:} with independent data and explicit characterization of uncertainty and bias.
  \item \textbf{Make governed decisions:} using prespecified rules and human review.
  \item \textbf{Monitor and revalidate:} tracking drift and model updates with version control and an audit trail, and returning to step~1 as the use evolves.
\end{enumerate}

\section{Discussion}\label{sec:discussion}

The five-level framework is intended as a shared vocabulary for statisticians, clinicians, data scientists, and regulators. Its main practical implication is that ``digital twin'' should never be evaluated as a single technology. A model that is entirely appropriate as a Level~1 covariate may be inappropriate as a Level~3 replacement control, even with identical predictive performance, because the inferential weight it carries differs.

Several implications follow for statisticians in industry. First, start where precedent exists: Level~1 prognostic adjustment fits within established covariate-adjustment practice and offers efficiency gains without compromising randomization-based validity. Second, treat every generated quantity as uncertain and propagate that uncertainty into the decision, whether through multiple draws, sensitivity analysis, or simulation. Third, insist on estimand alignment between training data and the current trial. Fourth, lead the prespecification, back-testing, and lifecycle governance that turn a prediction model into decision-grade evidence.

The framework has limitations. Positions in Figure~\ref{fig:credibility} are illustrative and depend on the specific context of use. The levels are not exhaustive; hybrid applications, such as virtual controls built on mechanistic components, may span categories. Quantitative examples cited here, including the AWARE analysis, are retrospective and should not be generalized without validation in the intended setting.

\section{Conclusion}\label{sec:conclusion}

Data-driven digital twins offer a spectrum of capabilities, from sharpening randomized comparisons to rehearsing entire trials in silico. Credibility requirements should scale with the model's influence on a decision and the consequences of that decision. If the field addresses the bottlenecks in data, model transferability, and governance together, and earns trust use by use, clinical development can become a continuously learning system. The goal is not synthetic data for its own sake, but better-informed development decisions that bring effective medicines to patients faster.

\section*{Disclaimer}
The views expressed are those of the author. This publication was neither originated nor managed by AbbVie and does not communicate results of AbbVie-sponsored scientific research.

\bibliographystyle{unsrtnat}
\bibliography{references}

\end{document}